\documentclass{natureprintstyle}

\usepackage{hyperref}

\usepackage{graphicx}%
\usepackage{multirow}%
\usepackage{amsmath,amssymb,amsfonts}%
\usepackage{amsthm}%
\usepackage{mathrsfs}%
\usepackage[title]{appendix}%
\usepackage{xcolor}%
\usepackage{textcomp}%
\usepackage{manyfoot}%
\usepackage{booktabs}%
\usepackage{algorithm}%
\usepackage{algorithmicx}%
\usepackage{algpseudocode}%
\usepackage{listings}%
\usepackage{rotating}
\theoremstyle{thmstyleone}%
\theoremstyle{thmstyletwo}%

\theoremstyle{thmstylethree}%

\newcommand{\gal}{Galaxies}    

\def\til{\ensuremath{\sim\,}}

\newcommand{\tim}[1]{\ensuremath{\times 10^{#1}}}
\def\deg{\ensuremath{^{\circ}}}

\def\msol{\ensuremath{M_\odot}}

\def\cms{\ensuremath{$cm$^{-2}}}

\def\swift{\emph{Swift}}

\def\nh{\ensuremath{N_{\rm H}}}

\def\t0{\ensuremath{T_{0}}}

\def\arcsec{\ensuremath{^{\prime\prime}}}
\def\arcmin{\ensuremath{^\prime}}
\def\nu{\ensuremath{\nu}}

\def\ergs{\ensuremath{$erg s$^{-1}}}

\def\s{\ensuremath{\sigma}}
\def\swj{Swift J0230}
\def\gal{2MASX J02301709+2836050}

\usepackage{longtable}
\usepackage[labelsep=endash]{caption}

\title{Monthly quasi-periodic eruptions from repeated stellar disruption by a massive black hole}

\author{P.A. Evans$^{1}$
{C.J. Nixon}$^{2,1}$
{S. Campana}$^{3}$
{P. Charalampopoulos}$^{4,5}$
{D. A. Perley}$^{6}$
{A.A. Breeveld}$^{7}$
{K.L. Page}$^{1}$
{S.R. Oates}$^{8}$
{R.A.J. Eyles-Ferris}$^{1}$
{D.B. Malesani}$^{9,10,11}$
{L. Izzo}$^{11,12}$
{M.R. Goad}$^{1}$
{P.T. O'Brien}$^{1}$
{J.P. Osborne}$^{1}$
{B. Sbarufatti}$^{3}$}

\begin{document}
\maketitle

\begin{abstract}
In recent years, searches of archival X-ray data have revealed galaxies
exhibiting nuclear quasi-periodic eruptions with periods of several hours. These
are reminiscent of the tidal disruption of a star by a supermassive black hole,
and the repeated, partial stripping of a white dwarf in an eccentric orbit
around a $\sim10^5 \msol$ black hole provides an attractive model. A separate
class of periodic nuclear transients, with significantly longer timescales,
have recently been discovered optically, and may arise from the partial
stripping of a main-sequence star by a $\sim10^7\ \msol$ black hole. No clear
connection between these classes has been made. We present the discovery of an
X-ray nuclear transient which shows quasi-periodic outbursts with a period of
weeks. We discuss possible origins for the emission, and propose that this
system bridges the two existing classes outlined above. This discovery was made
possible by the rapid identification, dissemination and follow up of an X-ray
transient found by the new live \swift-XRT transient detector, demonstrating the
importance of low-latency, sensitive searches for X-ray transients.
\end{abstract}

\let\thefootnote\relax\footnote{
\begin{affiliations}
\item{School of Physics and Astronomy, University of Leicester, University Road, Leicester, LE1 7RH, UK; pae9@leicester.ac.uk}
\item{School of Physics and Astronomy, University of Leeds, Woodhouse Ln, Leeds, LS2 9JT, UK}
\item{INAF, Osservatorio Astronomico di Brera, via E. Bianchi 46, 23807, Merate, Italy}
\item{Department of Physics and Astronomy, Tuorla Observatory, University of Turku, FI-20014, Turku Finland}
\item{DTU Space, Technical University of Denmark, Elektrovej 327, 2800, Kongens Lyngby, Denmark}
\item{Astrophysics Research Institute, Liverpool John Moores University, IC2, Liverpool Science Park, 146 Brownlow Hill, L3 5RF, Liverpool, UK}
\item{Mullard Space Science Laboratory, University College London, Holmbury St Mary, RH5 6NT, Dorking, UK}
\item{School of Physics and Astronomy \& Institute for Gravitational Wave Astronomy, University of Birmingham, B15 2TT, Birmingham, UK}
\item{Department of Astrophysics/IMAPP, Radboud University, 6525 AJ Nijmegen, The Netherlands}
\item{Cosmic Dawn Center (DAWN), Denmark}
\item{Niels Bohr Institute, University of Copenhagen, Jagtvej 128, 2200, Copenhagen N, Denmark}
\item{INAF-Osservatorio Astronomico di Capodimonte, Salita Moiariello 16, 80131, Napoli, Italy}
\end{affiliations}
}

\section{Introduction}

\begin{figure}
  \begin{center}
  \includegraphics[width=8.1cm,angle=0]{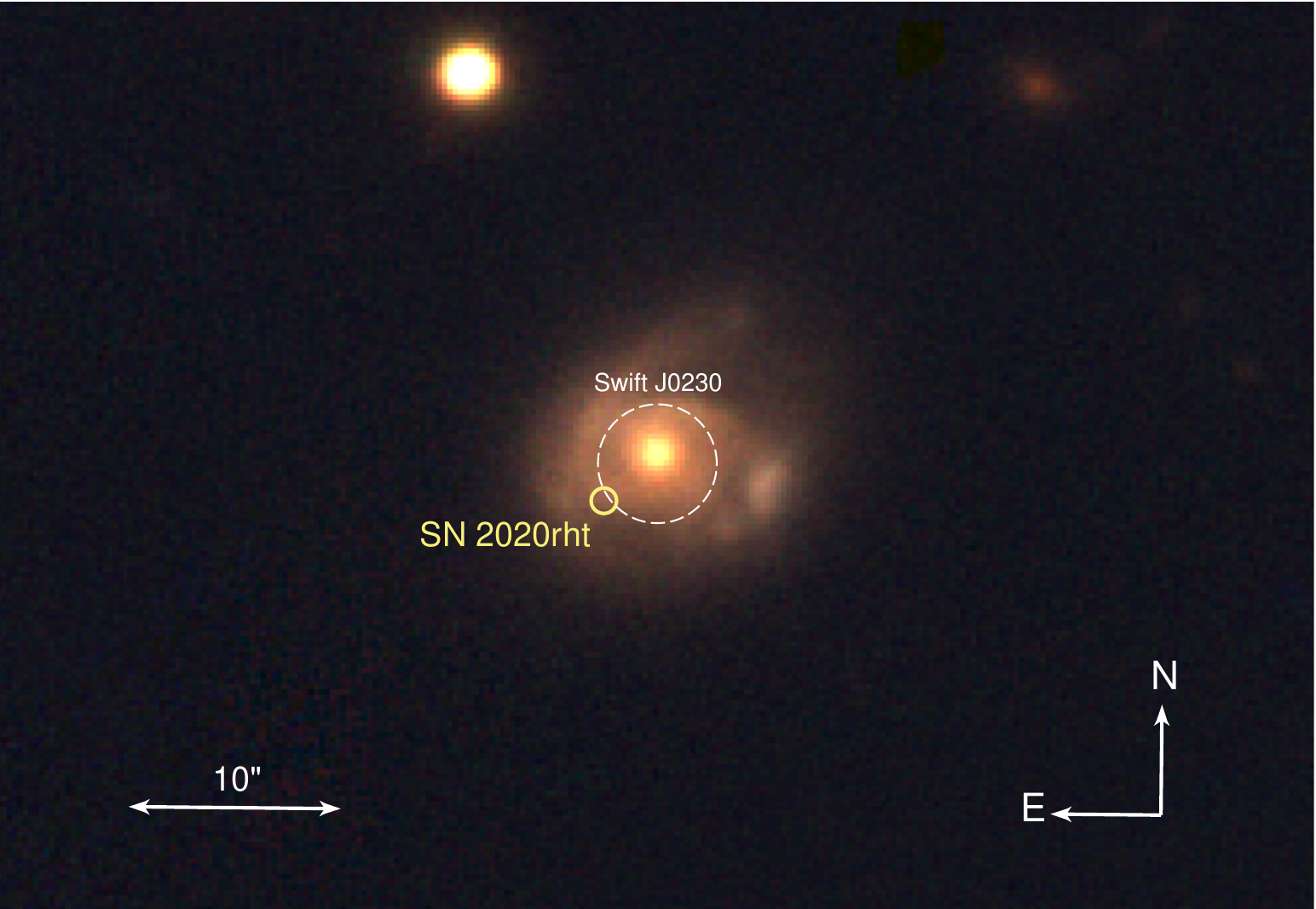}
  \end{center}
  \caption{The location of the new transient, \swj, relative to its host galaxy and an old supernova.
  The image is an archival Pan-STARRS \protect{\cite{PanSTARRS}} image of \gal, with colour scaled
  abitrarily for aesthetic purposes. The broken circle shows the 90\%\ confidence
  \swift-XRT position of \swj; the solid one SN 2020rht.
  }
  \label{fig:optical}
\end{figure}

Swift J023017.0+283603 (hereafter \swj) was discovered in \swift-X-ray Telescope (XRT) data by the
Living \swift-XRT Point Source (LSXPS) catalogue's real-time transient detector \cite{Evans23} on
2022 June 22 \cite{ATEL15454}. The source was serendipitously present in an observation of an
unconnected source, SN 2021afk (4.3\arcmin\ away), and had a 0.3--10 keV count rate of
$2.7^{+0.6}_{-0.5}\tim{-2}$ ct s$^{-1}$. This field had been observed on 11 previous occasions by
\swift\ between December 2021 and January 2022; combining all of those observations, \swj\ was
undetected down to a 3-\s\ upper limit of 1.5\tim{-3} ct s$^{-1}$. The last of these observations
was 164 days before the discovery of \swj, placing a rather loose lower limit on its switch-on time;
for convenience we give all times relative to MJD 59752 (midnight on the day of discovery). The best
localization of \swj\ is from the XRT, and is RA=$02^{\rm h}\ 30^{\rm m}\ 17.12^{\rm s}$, Dec=$+28\deg\ 36\arcmin\
04.4\arcsec$ (J2000), with an uncertainty of 2.8\arcsec\ (radius, 90\%\ confidence). This is
consistent with the nucleus of the galaxy \gal, but also marginally consistent with the type-II
supernova SN2020rht (3.1\arcsec\ away), discovered two years earlier on 2020 August 12
\cite{2020TNSTR2472....1T} (Figure~\ref{fig:optical}). An optical spectrum of \gal, obtained with
the Nordic Optical Telescope (Figure~\ref{fig:opt_spec}), gives a redshift $z=0.03657 \pm 0.00002$.
Assuming standard cosmological parameters \cite{Planck20}, this corresponds to a luminosity distance
$D_L=160.7$ Mpc. The galaxy type is unclear, but it is either quiescent or, at most, a very weak AGN
(see the Supplementary Information).

\section{Results}

\subsection{X-ray analysis}

Following the initial, serendipitous discovery, we obtained regular monitoring with \swift\ (see the
Methods section for details). The initial outburst continued for 4 days following
the discovery; on the fourth day it ended with a rapid decay, the luminosity falling by a factor of
20 in just 57 ks; there was a brief rebrightening (a factor of 4.5 in 6 ks), before it became too
faint to detect; the light curve is shown in Figure~\ref{fig:xrtlc}. Fitting this decay with a
power-law, $L\propto (t-t_0)^{-\alpha}$ (where $t_0$ is set to the start of the first bin in this
observation), gives $\alpha=11.0\pm 1.7$. Eight subsequent outbursts were observed at \til25~d intervals,
with durations \til10--15~d. The fifth outburst was either significantly longer (up to \til32~d), or consisted
of a weak outburst, a return to quiescence, and then a second, longer outburst. This was followed
by a long gap of \til 70~d during which two possible short and weak outbursts were seen,
before another outburst similar to the early ones.

A Lomb-Scargle analysis (see Methods and Extended Data Figure.~\ref{fig:ls}) reveals moderately-significant
peaks at approximately 22 and 25 d periods, although each peak is \til1 d wide, confirming the quasi- rather than
strictly-periodic nature of the variability, as may be expected from eyeballing the light curve.
Further consideration of the variability requires us to define what constitutes an outburst: the
initial outburst and that from days \til41--48 appear clearly defined, but during the outburst from
days \til60--75, the source underwent a sudden decline, being undetected on day 72 with an upper
limit of $L<8.9\tim{41}$ \ergs (0.3--2 keV), recovering to $L\til2\tim{42}$ \ergs\ by day 74. It
seems plausible to interpret this as a single outburst, with a sudden, brief dip. Hereafter, what
constitutes an outburst becomes more subjective. The outbursts starting on days 89 and 102 could
each be explained as comprising two short outbursts close together, or a single outburst with a
quiet phase in the middle; it is worth noting that in the first of these, if we sum the upper limits
during this quiet phase, we find a detection at a higher level than the upper limits found in the
quiescent phases. During the long, largely quiescent period from days 111--195 the source was twice
briefly detected with $L\til1-2\tim{41}$ \ergs, but these are hardly `outbursts' in the same way as
the earlier emission. Based on visual inspection of the light curve, we define an outburst as
comprising any times where $L>2\tim{41}$ \ergs; the details of the outbursts thus identified are
given in Table~\ref{tab:timescales} -- see the Methods section for full details of how these were
derived. In summary: we have detected transient X-ray emission that rapidly switches on and off
again with a recurrence timescale that is of the order of 25\,d, but which can vary by several days
between outbursts. The duration of the outbursts also shows significant variability with the longest
being of order 20 days and the shortest less than a day.

The X-ray spectrum during the outbursts was very soft, with no emission seen above 2 keV, and could
be well-modelled with a simple blackbody emitter with only Galactic absorption. Due to this
soft spectrum, the typical energy bands used for XRT hardness ratios were inappropriate; we selected
0.3--0.9 keV and 0.9--2 keV as this gave roughly equal counts in the two bands, maximising the
signal-to-noise ratio. The time-evolution of this hardness ratio (Figure~\ref{fig:xrtlc}) shows a
clear correlation between luminosity and spectral hardness (Spearman rank p-value of $1.3\tim{-6}$ of the
data being uncorrelated), ruling out a change in absorption as the cause of the flux variation. We
fitted the absorbed blackbody model to each observation in which \swj\ was detected. The blackbody
temperature obtained is strongly correlated with luminosity (p-value: $4.5\tim{-6}$; see Extended Data
Figure.~\ref{fig:lkt}); no evidence for absorption beyond the Galactic column is seen.

As noted earlier, while coincident with the galaxy nucleus, the XRT position is also potentially in
agreement with that of SN 2020rht. We obtained a 3 ks \emph{Chandra} DDT observation during the
fourth outburst to obtain a better position, but unfortunately this observation fell on day 97,
which turned out to be in one of the mid-outburst quiet phases.

\subsection{Optical and UV analysis}

At optical and ultraviolet wavelengths, there is no evidence for outbursting behaviour. We obtained data from both
\swift\ UV/Optical Telscope (UVOT) and the Liverpool Telescope (Extended Data Tables 1--2). The host galaxy, \gal, is clearly detected in all
observations, but there is no evidence for variability or an increase in flux compared to catalogued
values. For UVOT, we also analysed the data from the pre-discovery observations and find no secure
evidence for a change in brightness between those data and the observations taken during outburst.
Full details are given in the Methods section.

\section{Discussion}

The peak luminosity of $\til4\tim{42}$ \ergs, the timescales of the outbusts and their
quasi-periodic, quasi-chaotic nature, the soft X-ray spectrum and lack of optical variability place
significant constraints on the possible models to explain \swj. While the lack of detection with
\emph{Chandra} means that we cannot rule out a positional association with SN2020rht, it is
difficult to see how a supernova could have evolved into the object which we have detected. The spectrum,
luminosity and variability timescale are inconsistent with the properties of ultra-luminous X-ray
sources \cite{Kaaret17}, and while certain supernovae could in principle be followed by X-rays from
a newly-formed millisecond pulsar \cite{Metzger14}, this should occur while the supernova is still
visible in the optical. \cite{Margalit18} have shown how this emission could be delayed, but the timescales
and luminosities they predict (e.g. their fig.~5) are not consistent with our observation. Equally, neither model
explains the variability or spectrum we see in \swj. We discuss this further in the Supplementary Information.

We suggest that (near) periodic mass supply into an accretion flow onto the central supermassive
black hole in \gal\ presents the most likely explanation for \swj. From simple energetics (see
Supplementary Information) we can infer that the total mass accreted during a typical outburst is
$\til10^{-5}\ \msol$. In an AGN, a supermassive black hole at the heart of a galaxy accreting from a
surrounding disc of gas, flares and outbursts are common. However, the timescale and spectrum of
\swj\ are not consistent with typical AGN behaviour, and \gal\ itself does not appear to be an AGN
(Extended Data Figure.~\ref{fig:bpt}; see Supplementary Information for a full discussion). We therefore
consider the possibility that one (or several) stars are interacting with, and feeding mass on to,
the central supermassive black hole. One possible mechanism for producing the mass flow is the
interaction of two stars in orbit around the black hole; if they pass sufficiently close to each
other material can be liberated from one or both stars that can feed the central black hole
\cite{Metzger17b,Metzger22}. To generate the required timescales from this model requires a pair of
stars orbiting in the same direction and in the same plane \cite{Metzger22}. This could occur for
stellar orbits that are initially randomly oriented if they can be subsequently ground down into the
same plane by interaction with an AGN disc \cite{Syer91}. For \swj, which lacks any clear signature
of a standard AGN disc, it is unlikely that any stars orbiting the central black hole have the
required orbits to achieve the observed timescales.

Another possibility is a repeating, partial Tidal Disruption Event (rpTDE), in which a star on a
bound, highly-eccentric orbit loses some of its envelope every pericentre passage due to tides from
the black hole's gravitational field. These events are a sub-class of TDEs in which the ``regular'' scenario sees the incoming star
approach the black hole on a parabolic orbit, and the star is destroyed by the first encounter (see refs~\cite{Rees88,Gezari21} for reviews of
TDEs).

The rpTDE model was investigated prior to the discovery of any corresponding sources (for example, ref~\cite{Zalamea10})
 and was suggested as the explanation of X-ray flares in the active galaxy
IC 3599 \cite{Campana15}. Recently, it has been proposed (for example, refs~\cite{King20,King22,Lu23}) as a
possible explanation for hours-long quasi-periodic eruptions (QPEs) discovered in galactic nuclei
(for example, refs~\cite{Miniutti19,Giustini20,Song20,Chakraborty21,Arcodia21}). These works focused on the
possibility of a white dwarf interacting with a relatively low-mass central black hole of mass $\til
10^{5-6} \msol$. A second set of sources show much longer outbursts (both in duration and recurrence
period) and have been referred to as periodic nuclear transients (PNTs); these may be the same rpTDE
mechanism acting with a main-sequence star rather than a compact star
\cite{Payne21,Payne22,Wevers22,Liu23}, and a more massive black hole ($\til 10^{7-8} \msol$).

In the rpTDE model, the donor star is in a highly eccentric orbit around a black hole; at each
pericentre passage the star has to approach, but not quite reach, the tidal radius at which the star
would be fully disrupted. The outer layers of the star are liberated, and some of this material
accretes on to the central black hole powering the outburst. The recurrence time of the outbursts
is related to the orbital period of the star. The majority of the energy released from the
accretion process occurs in the central regions near the black hole where the matter is most likely
in the form of an accretion disc. We can therefore provide an estimate of the black hole mass in
\swj\ by comparing the temperature of $\til 100$\,eV ($\sim 10^6\,$K), measured from the X-ray
spectrum, with the peak temperature of a standard accretion disc \cite{Shakura73}. This yields (see
the Supplementary Information) a black hole mass estimate of $\sim 2\times 10^{5}\msol$. This is similar to
the mass estimates for the QPE sources (for example, refs~\cite{Miniutti19,Giustini20,Arcodia21}). It is worth
noting that the QPE sources and \swj\ show very little in the way of optical emission, whereas the
PNTs show strong optical emission. This may simply reflect the difference in the black hole masses,
i.e.\ the soft X-ray spectrum and the lack of optical emission seen in \swj\ appears consistent with
this estimate of the black hole mass.

If the accreted material is stripped from the orbiting star during pericentre passage of a highly
elliptical orbit, and the pericentre distance is a few gravitational radii (required to liberate any
material from the surface of a white dwarf for black hole masses of a few $\times10^5\,\msol$), then
we would expect the outburst duration to be similar in different systems, regardless of their
orbital (and hence outburst) period. This is because the pericentre passage of a highly elliptical
orbit is approximately that of a parabolic orbit and its duration is not connected to the orbital
period. This means that it is difficult to explain both \swj\ (outburst duration of days) and the
pre-existing QPEs (outburst duration of hours) as rpTDE of a white dwarf around a modest-mass black
hole. On the other hand, rpTDE of main-sequence stars by $\til 10^7$ -- $10^8\ \msol$ black holes
have been proposed to explain the PNTs ASASSN-14ko, for which the recurrence timescale is 114~d
\cite{Payne21,Payne22}, and AT2018fyk which exhibited a significant re-brightening after around
600~d of quiescence \cite{Wevers22}. \swj\ clearly lies between these two classes of object. 

An important question is how the star arrived on such an orbit around the central black hole. Tidal
capture, in which an orbiting star loses orbital energy due to tidal forces and becomes bound to the
black hole \cite{Fabian75}, is typically incapable of generating the required orbits; however, the
Hills mechanism \cite{Hills98} was proposed as a viable formation route for the PNT ASASSN-14ko
\cite{Cufari22}. In this mechanism, a binary star system approaches the black hole with a small
enough pericentre distance such that the tidal force from the black hole is stronger than the
gravitational force holding the binary together. This results in the binary being disrupted, with
one component ejected from the system, and the other locked into a bound, but highly eccentric,
orbit about the black hole. If the progenitor of \swj\ were a binary consisting of a low mass,
main-sequence star and, say, a white dwarf, then the main-sequence star needs to be captured into a
bound orbit around the black hole with the observed \til25 d period. For a black hole mass of
$M_\bullet \sim 4\times 10^5\,\msol$ this period corresponds to the most likely outcome of Hills
capture from such a binary system (the calculations of \cite{Cufari22} show that higher black hole
masses are allowed but are significantly less likely to result in this period for the bound star;
see Supplementary Information for details). This is consistent with the mass estimate derived from the temperature of
the X-ray spectrum. Further, the expected accretion timescales from such a system (see Supplementary Information) are
also consistent with those observed in \swj.

The variable shape and timescales of the outbursts seen in \swj\ may also be explained by
this model. In a standard TDE, as opposed to an rpTDE, the star arrives on a parabolic orbit,
meaning that some of the stellar matter is bound to the black hole (the inner tidal stream), and
this material forms an accretion flow, while the rest of the stellar debris (the outer tidal stream)
is unbound and leaves the system \cite{Rees88}. In an rpTDE, the star must be on a bound orbit
around the black hole. In this case both the inner and outer tidal streams can remain bound to the
black hole. The inner stream falls back soonest, and thus with a higher mass return rate, while the outer stream can return on longer timescales. Due to relativistic
precession of the stellar debris orbits (both apsidal and nodal) the returning streams can collide
and partially cancel their angular momenta to augment the accretion rate on to the black hole, with
the magnitude of the effect depending on the orientation of the colliding orbits (see
\cite{Nixon12,Raj21} for similar variability induced in accretion discs due to relativistic
precession; and the processes described therein may also occur in the discs formed in \swj). The exact details of this interaction between the two streams, the accretion flow, and
the orbiting star are complex and require a full numerical analysis which is beyond the scope of
this discovery paper; however it is clear that such interaction will produce variable emission that
could at least partially erase the more exactly periodic nature of the stellar orbit. An example of such complex
interacting debris streams can be seen in fig.~9 of \cite{Coughlin17}. It will be particularly important to determine if the star itself can be sufficiently perturbed during each pericentre passage, with e.g. tides imparting variations in oscillation amplitudes and rotation frequency, to change the amount of mass transferred and the structure of each outburst. Additionally, the sharp decline observed at the end of each outburst may be driven by the returning star disturbing the accretion flow. These questions can be addressed with future theoretical investigations.

We have proposed a single explanation for the QPEs and PNTs, as repeated, partial tidal disruption
of a star in an eccentric orbit around a supermassive black hole; and reported the discovery of the
first object that can bridge the gap between these classes. The QPEs are thought to harbour a white
dwarf and a modest ($\til 10^5$ \msol) black hole, and the PNTs are thought to host a main sequence
star and a more massive ($\til 10^7$ \msol) black hole. \swj\ represents an intermediate class of
system which is consistent with a main sequence star orbiting a modest-mass black hole. Given the
timescales, modest fluxes, and lack of emission outside of the X-ray band, \swj-like systems are
difficult to discover. Unlike QPEs, which were discovered in archival data, their timescales and
behaviour would not be exposed by a single observation. It is only with the recent creation of a
real-time transient detector \cite{Evans23} that objects like this can be found rapidly enough for
follow-up observations to expose their behaviour. The fact that this event was found within three months
of enabling this real-time search suggests that they are reasonably common and we can expect to
discover more objects of this class with sensitive, wide-field X-ray instruments such as
\emph{eRosita} \cite{Predehl21} and in the near future, the \emph{Einstein Probe} \cite{Yuan22}.

\begin{figure}
  \begin{center}
  \includegraphics[width=8.1cm,angle=0]{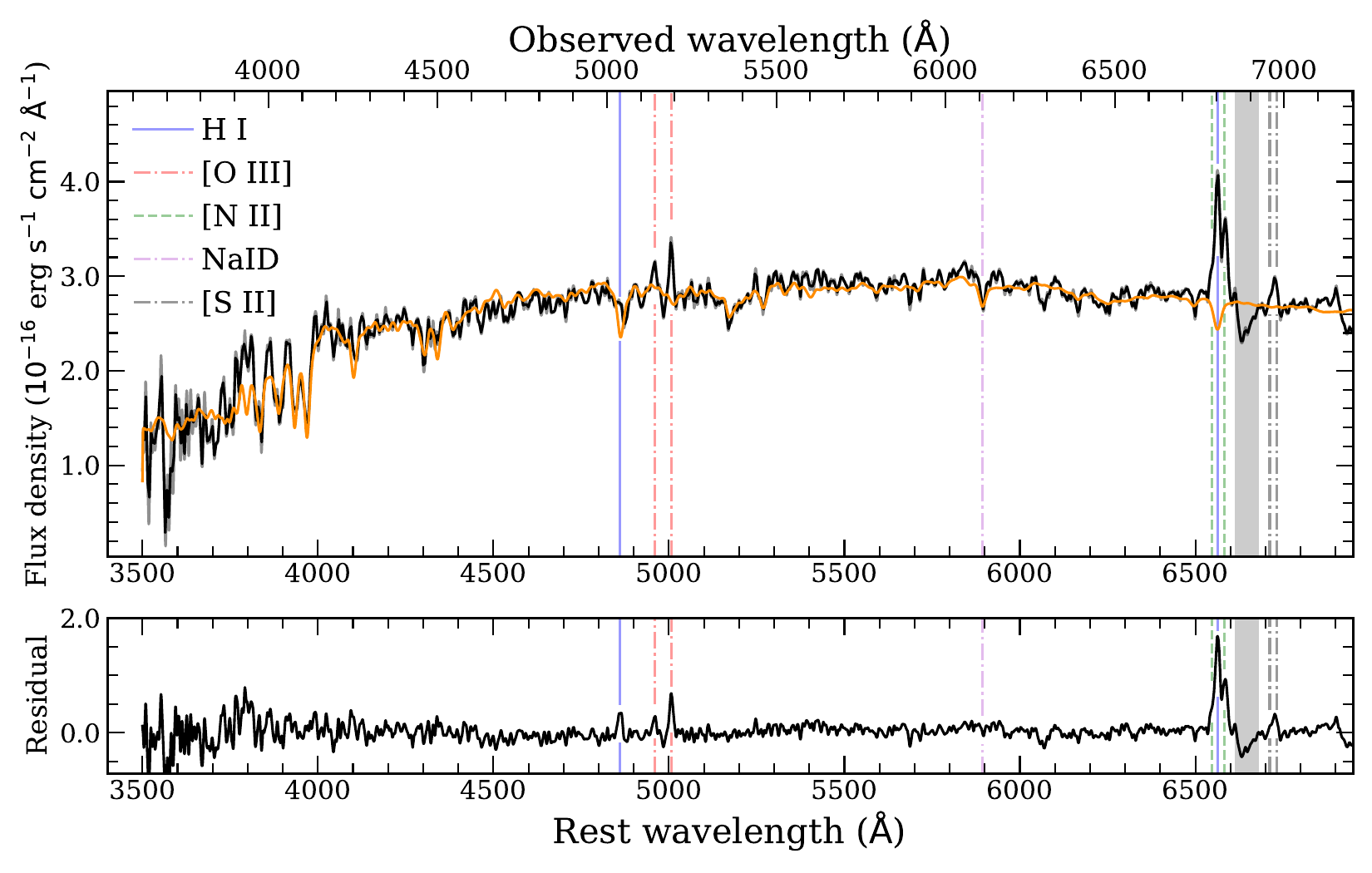}
  \end{center}
  \caption{Optical spectrum of the host galaxy \gal\ obtained
with the Nordic Optical Telescope on day 132. \emph{Top}: the
black line shows the observed spectrum, while the orange line shows the
fit to the stellar continuum provided by {\sc starlight}. The
vertical lines mark prominent emission and absorption features, which
together allow to measure the redshift $z = 0.03657 \pm 0.00002$ ($1-\sigma$ confidence). \emph{Bottom}:
the residuals between the observed data (stellar + nebular
spectrum) and the fit (stellar continuum), which single out the nebular
emission. The emission line fluxes were measured on the residual
spectrum and allow placing the galaxy on the BPT plot (Extended Data
Figure.~\ref{fig:bpt}). }
  \label{fig:opt_spec}
\end{figure}

\begin{figure*}
  \begin{center}
  \includegraphics[width=16cm,angle=0]{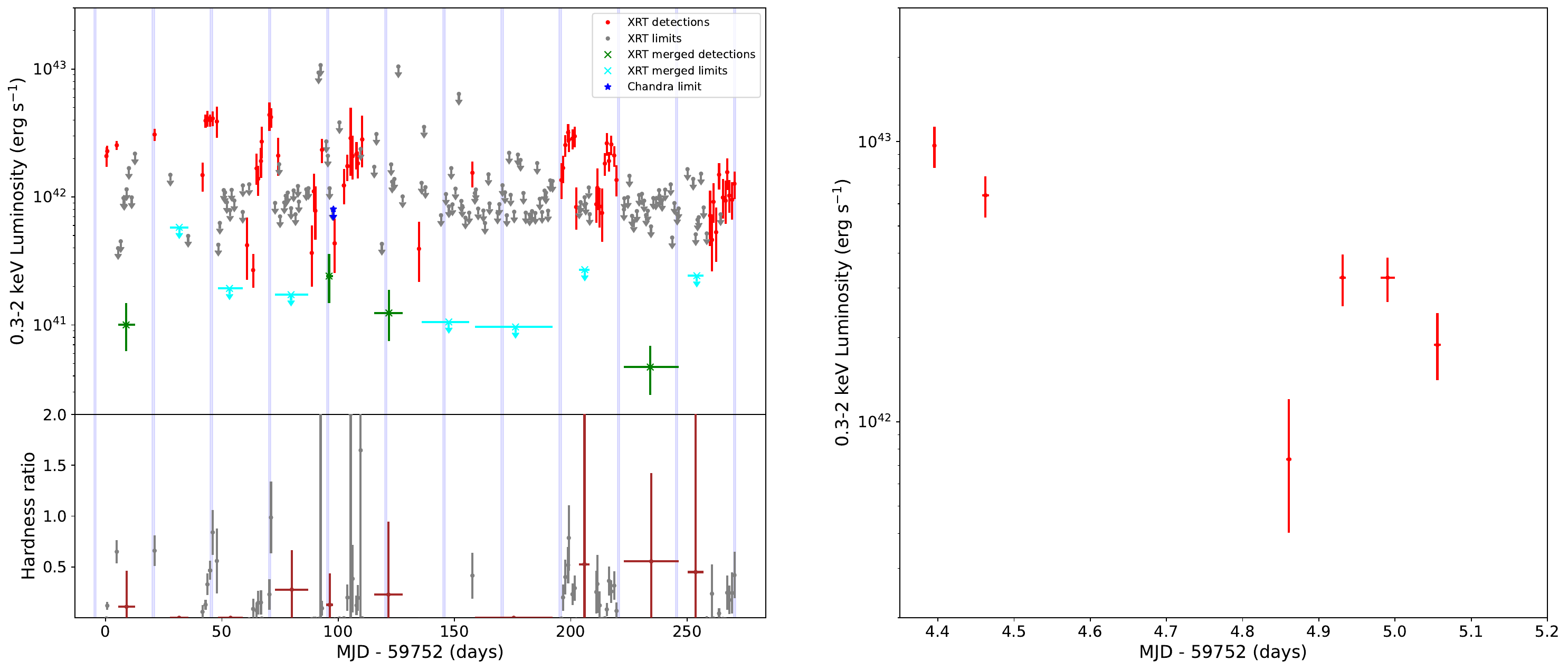}
  \end{center}
  \caption{The temporal evolution of \swj. All error-bars are $1 \sigma$ significance. {\bf Left:} X-ray time series, binned to one bin per observation. \emph{Top}: 0.3--2 keV
  luminosity light curve. The red data points marked with dots are the \swift-XRT detections; grey
  arrows are 3-\s\ upper limits from XRT. The dark blue upper limit marked with an asterisk is from
  \emph{Chandra}. The broad bins marked with crosses were created by combining consecutive XRT
  non-detections (upper limits in cyan, detections in green); see the Methods section for full
  details. \emph{Bottom}: The
  (0.9--2)/(0.3--0.9) keV hardness ratio; the spectral hardness is strongly correlated with the
  luminosity. The vertical bands are at 25-d intervals.
  {\bf Right:} the light curve of the XRT observation taken on days 4--5, with one bin per
  spacecraft orbit, showing the rapid decay at the end of the first outburst.
  All error-bars are at the $1 \sigma$ level.}
  \label{fig:xrtlc}
\end{figure*}

\begin{sidewaystable}
  \begin{center}
  \begin{minipage}{\textheight}
  \caption{Constraints on the timescales of the observed outbursts. $^*$The final outburst was ongoing when observations
  ended.}\label{tab:timescales}
  \begin{tabular*}{\textheight}{@{\extracolsep{\fill}}ccccccccc@{\extracolsep{\fill}}}
    \hline
      Outburst & \multicolumn{2}{c}{Start}    & \multicolumn{2}{c}{End}        & \multicolumn{2}{c}{Duration}    & \multicolumn{2}{c}{Period} \\
           & MJD min  & MJD max           & MJD min & MJD max              &  min (d)   &  max(d)            &  min (d)    &  max(d)  \\
  \midrule
  1 & 59587.9  & 59752.3                  & 59757.3 & 59757.4              & 4.9        & 169.5 & \\
  2 & 59763.2  & 59773.0                  & 59773.1 & 59787.5              & 0.1        & 24.3               & 10.9   & 185.1  \\ 
  3 & 59787.5  & 59793.7                  & 59799.8 & 59800.4              & 6.2        & 12.9               & 14.5   & 30.4   \\
  4 & 59811.0  & 59815.3                  & 59826.4 & 59827.0              & 11.1       & 16.0               & 17.3   & 27.8   \\
  5 & 59839.2  & 59841.5                  & 59862.3 & 59867.5              & 20.8       & 28.3               & 24.0   & 30.5   \\
  6 & 59870.8  & 59886.7                  & 59886.7 & 59887.9              & 0.0        & 17.0               & 29.3   & 47.5 \\ 
  7 & 59908.3  & 59909.7                  & 59909.7 & 59910.8              & 0.0        & 2.4                & 21.6   & 38.9 \\ 
  8 & 59942.3  & 59948.0                  & 59954.3 & 59955.6              & 6.4        & 13.4               & 32.5   & 39.6 \\ 
  9 & 59960.2  & 59963.0                  & 59971.7 & 59974.8              & 8.7        & 14.7               & 12.2   & 20.7 \\ 
  10 & 60010.8 & 60011.9                  & 60022.4* & ---                 & 11.6*      & ---                & 47.8   & 51.8 \\ 
  \hline
  \end{tabular*}
  \end{minipage}
  \end{center}
  \end{sidewaystable}

\renewcommand{\figurename}{Extended Data Figure} 
\renewcommand{\tablename}{Extended Data Table} 
\setcounter{figure}{0}
\setcounter{table}{0}

\section*{Methods}\label{sec11}

\section*{Discovery}

At 13:58 UTC on 2022 June 22, the LSXPS real-time transient detector \cite{Evans23} reported the
discovery of a possible new X-ray transient, dubbed Swift J023017.0+283603. The object was detected
in \swift\ observation 00014936012 which had taken place between 08:19 and 08:46 UTC; i.e. the
notification was produced 5.2 hours after the observation (most of this latency related to
the timing of ground station passes and the ingesting of the data by the Swift Data Center: the
observation data were received at the UKSSDC at 13:51 UTC). No catalogued X-ray source was found at
this position. Further, \swift\ had already observed this location on 11 previous
occasions (the observation target was SN2021afkk, 4.3\arcmin\ away from this serendipitous
transient), for a total of 9.6 ks. These observations had been analysed as a stacked image in
LSXPS (DatasetID: 19690); no source was present near the position of \swj, with a
\hbox{3-\s}\ upper limit of 1.5\tim{-3} ct s$^{-1}$. The peak count rate of \swj\ in the new
observation was $2.7^{+0.6}_{-0.5}\tim{-2}$ ct s$^{-1}$ (1-\s\ errors), significantly above this
upper limit, clearly indicating that a new transient had been discovered.

Due to the very soft spectrum and coincidence with the nucleus of the galaxy \gal, this was
originally interpreted as a tidal disruption event \cite{ATEL15454,ATEL15461}, and a high-urgency
target-of-opportunity (ToO) request was submitted to \swift.

In the following analysis we assumed $H_0=67.36$ km s$^{-1}$ Mpc$^{-1}$, $\Omega_\Lambda=0.6847$,
$\Omega_m=0.3153$ \cite{Planck20}.

\section*{Observations and data analysis}

\swift\ follow-up observations began at 16:07 UTC on June 22 (T0+0.67 d). Due to \swift's Moon
observing constraint, subsequent observations were not available until day 4 (June 26). Daily
observations of 1 ks exposure were obtained with \swift\ until day 12. A subsequent ToO request was
submitted (PI: Guolo) requesting weekly monitoring of this source, which began on day 21 (2022 July
13). The initial observation showed that the source had turned back on in X-rays, but in the
following observations on days 27 and 35 it was again below the detection threshold. In order to
better quantify the duty cycle, we submitted regular ToO requests (PI: Evans) for daily 1 ks
observations, which ran until 2023 March 19 (day 270) when the source entered \swift's Sun observing
constraint. Note that we have not \emph{obtained} 1 ks per day; each month proximity to the Moon
prevents observations for 3--4 days, and due to the nature of \swift's observing programme, our
observations were sometimes shortened or completely superseded by other targets.

\subsection*{Swift-XRT}

XRT data were analysed using the on-demand tools of \cite{Evans09}, via the {\sc swiftools Python}
module (v3.0.7). A 0.3--10 keV light curve was constructed, binned to one bin per observation; the
soft and hard bands were set to 0.3--0.9 and 0.91--2 keV respectively. Observations 00015231018 and
00015231019 overlapped in time, as did 0001523143 and 0001523144. When this happens the
per-observation binning is unreliable, so we built light curves of each of these observations
individually, and then replaced the affected bins in the original light curves with those thus
obtained. For each run of consecutive upper limit bins, we merged the limits into a single bin, using {\sc
mergeLightCurveBins()} function in {\sc swifttools}, giving a better measurement of, or limit on,
the quiescent flux.

For each observation in which the light curve showed a detection (at the 3-\s\ level), we extracted a
spectrum, fitting it with a blackbody component absorbed by two absorbers. The first was a {\sc
tbabs} model with \nh\ fixed at the Galactic value of 1.12\tim{21} \cms\ \cite{Willingale13}; the
second was a {\sc ztbabs} model with \nh\ free to vary, and the redshift fixed at the value obtained
from our NOT spectrum. From these fits we obtained the 0.3--2 keV flux and (given the luminosity
distance of 160.7 Mpc) luminosity. The dependence of this luminosity on blackbody temperature,
reported in the main paper, is shown in Extended Data Figure.~\ref{fig:lkt}. We also obtained for each spectrum
the conversion factor from 0.3--10 keV count rate to 0.3--2 keV luminosity, and so converted our
count-rate light curve into luminosity. For the detections with too few photons to yield a spectral
fit, the upper limits in the light curve and the bins created by merging (above), we used the
conversion factor of $7.54\tim{43}$ erg ct$^{-1}$ obtained from the discovery observation. The
resultant light curve was shown in Figure~\ref{fig:xrtlc}, with the merged bins marked with crosses
(and in green/cyan in the online version). To explore the rapid flux decay seen at the end of the
first outburst (right-hand panel of Figure~\ref{fig:xrtlc}), we rebinned the data to one bin per snapshot
(using the {\sc rebinLightCurve()} function), converting to luminosity using the same conversion
factor (that obtained for the appropriate observation) for each bin.

In order to determine the timescales of the outbursts (Table~\ref{tab:timescales}), we defined
outbursts as being times where the 0.3--2 keV luminosity was above $3\tim{41}$ \ergs\ (based on
visual inspection of the light curve). We built a new light curve, still with one bin per
observation, but in which all bins were created as count-rates with 1-\s\ errors, rather than
allowing upper limits (using the {\sc swifttools} module with the argument {\sc allowUL=False}
passed to the light curve), and then identified each point which was inconsistent with $L=
3\tim{41}$ \ergs\ at at least the 1-\s\ level. The start and end times of the outbursts were then
constrained to being between consecutive datapoints from this sample which were on opposite sides of
the $3\tim{41}$ \ergs\ line. The results are shown in Table~\ref{tab:timescales}. We created a
Lomb-Scarle periodogram (using the {\sc astropy.timeseries.LombScargle Python} package) to search
this light curve for periodicity, and find possible peaks centred on 22.1 d and 25.0 d, each with
widths of \til1 d. To determine their significance we used a bootstrapping method, whereby one `shuffles' the
data, randomly redistributing the fluxes (and their errors) among the time bins, and then
recalculates the periodogram. We did this 10,000 times, and then for each trial period in the
periodogram identified the 99.7th percentile of power, i.e.\ the 3-$\sigma$ significance threshold.
The result is shown in Extended Data Figure.~\ref{fig:ls}, along with the window function. These two peaks
are both clearly above 3-$\sigma$ in significance and not present in the window function.

We also investigated whether with our observing strategy we can rule out short-period variations
like those seen in the QPEs reported to date. We simulated a simplistic light curve based on the
period of GSN 069 (and including alternating between slightly longer and shorter cycles as in GSN
069). For each snapshot in the real XRT light curve of \swj\ we determined the phase of the trial
period and set the count rate either to 0.03 ct s$^{-1}$ (`on') or 0.001 ct s$^{-1}$ (`off');
fractional errors were set to typical values from our real light curve. We then constructed the
Lomb-Scargle periodogram of this, and repeated the bootstrap approach above. A strong signal was
found at the nomimal period, confirming that such a signal would have been easily recovered had it
been present. Thus we can be confident that there is no short-period modulation like that in GSN 069
present in \swj.

\subsection*{Swift-UVOT}

UVOT data were analysed using the {\sc uvotsource} package in {\sc heasoft} v6.30. For the
pre-outburst data, the location of \swj\  was only in the field of view on five occasions. No sign
of variability is found in these data, so we summed the images in each filter using {\sc uvotimsum}
and extracted mean magnitudes. In the initial discovery observation UVOT gathered data in all
filters, but no sign of the outburst is seen, any variability being swamped by the underlying galaxy
emission; UVOT magnitudes from this observation and the pre-discovery data are shown in Extended
Data Table~\ref{tab:uvotPhot}. Due to this lack of variability, subsequent \swift\ observations used
the UVOT `filter of the day', which rotates each day between the $u$ and UV filters, to preserve the
life of the filter wheel. No significant variability is seen. When visually examining the light
curve it is tempting to claim some variability in phase with the XRT data, but the magnitude of the
variability is much smaller than the errors on the UVOT photometry. To further investigate, we
rebinned the XRT light curve to one bin per snapshot (i.e.\ the same binning as the UVOT data, which
has one exposure per snapshot), and disabled upper limits, forcing a count-rate and 1-\s\ error per
bin. For each UVOT filter, we identified the coincident XRT data and then performed a Spearman rank
correlation analysis between the XRT and UVOT fluxes. This does not account for the uncertainty on
the count-rates, and therefore is likely to overestimate the significance; however, no significant
correlation was found at all, with p-values between 0.1 and 0.9, and so more complex correlation
mechanisms were not deemed necessary. We also attempted image subtraction, summing all UVOT images
in the $u$ filter (that which showed the strongest signs of possible variability) during times of
XRT detection and non-detection, before subtracting the latter from the former. No evidence of an
excess at the XRT position was seen.

\subsection*{Nordic Optical Telescope}

Spectroscopy of the host galaxy \gal{} was obtained on 2022 November 1 (day 132). A 2.4-ks optical
spectrum was accumulated using the Alhambra Faint Object Spectrograph and Camera (ALFOSC) mounted on
the 2.56-m Nordic Optical Telescope (NOT) located at La Palma, Spain. The ALFOSC spectrum was
reduced using the spectroscopic data reduction pipeline PyNOT ({\url
{https://github.com/jkrogager/PyNOT}). We used a 1.0\arcsec\ slit width and Grism \#4, covering the
wavelength range $\sim3200$--9600~\AA\ at resolution $\Delta{\lambda}/\lambda \approx 360$. The
airmass during the observation was of the order of $\sim 1.1$. The spectrum is shown in Figure~\ref
{fig:opt_spec}, and features prominent H$\alpha$, [N\,\textsc{ii}] and [O\,\textsc{III}] emission
lines, with a common redshift of $0.03657 \pm 0.00002$. A weak H$\beta$ line is also seen. The flux
of weaker lines is often affected by the presence of stellar absorption in the continuum. In order
to recover the pure nebular fluxes, we fitted the spectrum with the {\sc starlight} (\url
{http://www.starlight.ufsc.br/}) software. {\sc starlight} fits the stellar continuum, identifying
the underlying stellar populations in terms of age and metallicity. Comparing the observed data with
the output synthetic spectrum, the pure nebular continuum can be identified, and the emission line
fluxes measured accurately.

Based on this analysis, we could build the `BPT' (Baldwin, Phillips \& Terlevich) diagram
\cite{Baldwin81}, which is widely adopted to identify the level of nuclear activity in a galaxy. It
exploits the ratio of nearby emission lines, minimizing the effects of extinction. The result is
shown in Extended Data Figure.~\ref{fig:bpt}. \gal{} lies in the locus where low-power AGN, LINERs (Low-Ionization
Nuclear Emission-line Region) and star-forming dominated galaxies intersect. A secure classification
of \gal\ is therefore not possible, but a powerful AGN is clearly ruled out.

\subsection*{Liverpool Telescope}

The position of \swj\ was observed with the 2\,m Liverpool Telescope (LT) optical imager (IO:O) on six
different occasions between 2022 Jun 28 and 2022 Aug 26, using the $griz$ filters (the first and
last epoch were $gri$ only). Images were processed using the default IO:O pipeline and downloaded
from the LT archive. We co-added exposures and performed basic image subtraction versus Pan-STARRS 1
reference imaging using a custom IDL routine. While a few subtracted image pairs show weak positive
or negative residuals at the location of the $\sim$18 mag nuclear point source, there is no clear
correlation in these residuals between filters or epochs, suggesting minimal contribution of any
nuclear transient (or any AGN variability) to the optical flux at the sensitivity level of the LT
images. The lack of any residual source at the location of SN\,2020rht is unambiguous in all images.
Limiting magnitudes of the images (5$\sigma$) are given in Extended Data Table~\ref{tab:ltPhot}.

\subsection*{Chandra}
We requested a 3 ks Director's Discretionary Time observation of \swj\ with \emph{Chandra} (Proposal
23708869). We triggered this on day 93, when the \swift-XRT count rate exceeded the approved
threshold of 0.02 ct s$^{-1}$, and the observations were obtained on day 97 (obsID 27470). Our
intention was to obtain an accurate (arc-second or better) position of \swj, to be able to say
definitively whether it was associated with the nucleus of its host, the historical supernova, or
neither. Unfortunately, this observation occured during the quiescent/faint part of the 5th
outburst, and \swj\ was not detected. The 3-\s\ upper limit, converted to 0.3--2 keV luminosity
assuming a 90 eV blackbody with a Galactic absorber, is $L<8.0\tim{41}$ \ergs, consistent with the XRT 
measurements at the time (Fig.~\ref{fig:xrtlc}).

\pagebreak

\begin{figure}
  \begin{center}
    \includegraphics[width=8.1cm,angle=0]{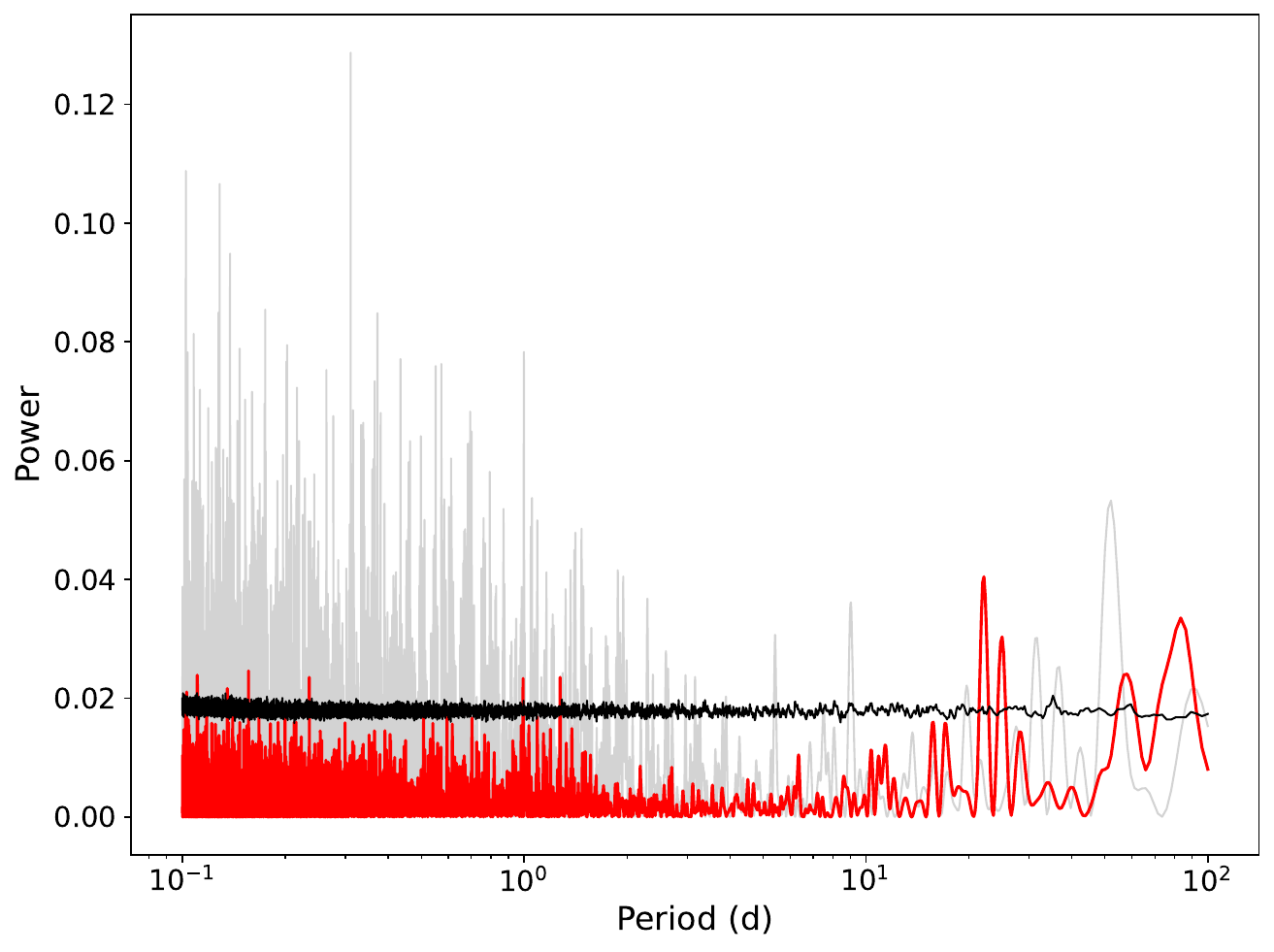}
  \end{center}
  \caption{Period analysis of the XRT data of \swj. The Lomb-Scargle periodogram of the per-snapshot binned XRT light curve
  is shown in red. The window function is in grey and the line marking 3-$\sigma$ significance as a function of period
  is in black. The two peaks above the 3-$\sigma$ line and not corresponding to window-function peaks are centred on 22.1 d and 25.0 d.}
  \label{fig:ls}
\end{figure}

\begin{figure}
  \begin{center}
  \includegraphics[width=8.1cm,angle=0]{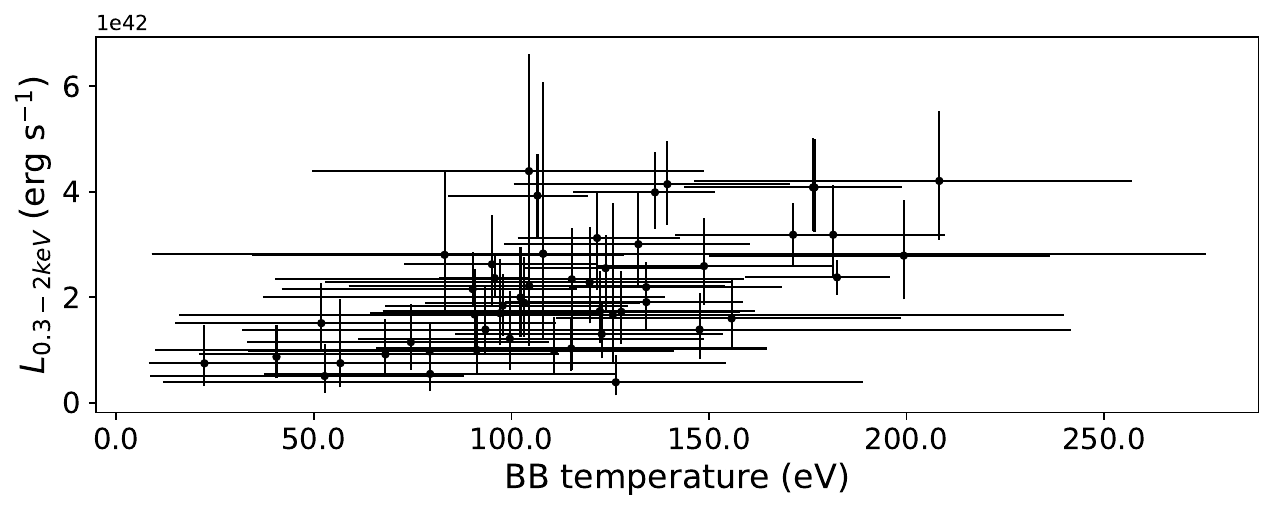}
  \end{center}
  \caption{The 0.3--2 keV luminosity and blackbody temperature derived from spectral fits to the XRT
  observations in which \swj\ was detected. A Spearman rank test shows these to be strongly correlated
  (p-value: $4.5\tim{-6}$). The errorbars reflect the 90\%\ confidence intervals on the parameters, obtained
  using $\Delta C = 2.706$ in the spectral fitting.
  }
  \label{fig:lkt}
\end{figure}

\begin{figure}
  \begin{center}
    \includegraphics[width=8.1cm,angle=0]{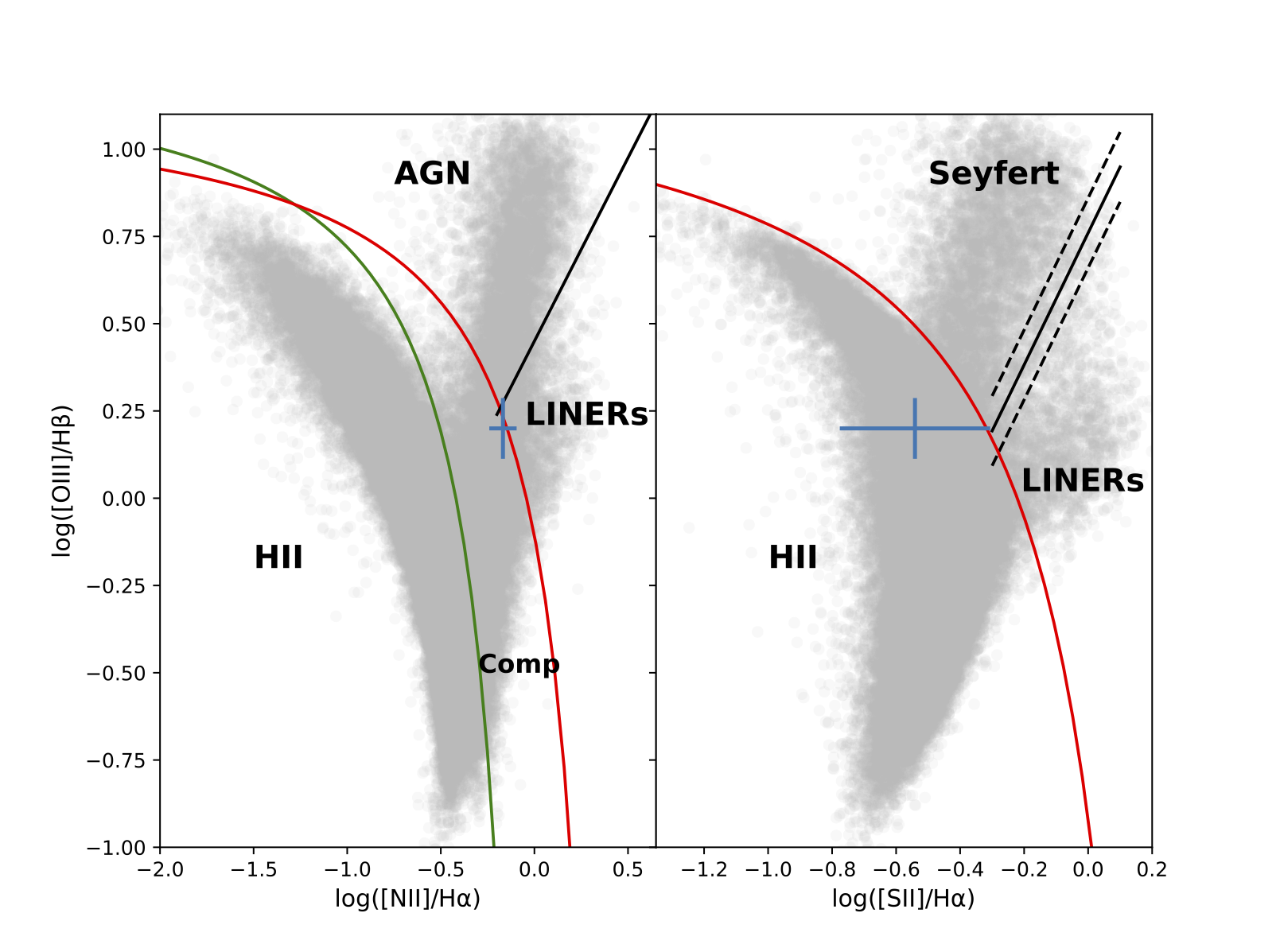}
  \end{center}
  \caption{BPT (Baldwin, Phillips \&\ Terlevich) diagram, showing galaxy type (H{\sc II} star forming region,
   AGN, LINER, composite) as a function of certain line flux ratios. The line ratios are [O\,\textsc{iii}] 5007 /
   H$\beta$ versus [N\,\textsc {ii}] 6583 / H$\alpha$ (left) and [S\,\textsc{ii}] 6717,6732 /
   H$\alpha$ (right). In both panels, the red solid lines are the theoretical models separating
   star-forming regions and AGN \cite{Kewley01}. In the left panel, the green line is the demarcation between
   pure star forming and composite star-forming/AGN regions, as prescribed by \cite{Kauffmann03}. 
   The straight segments separate proper AGN from LINERs (left:
   \cite{Schawinski07}; right: \cite{Kewley01}).
   The SDSS galaxy catalogue object
   density is shown in greyscale \cite{Tremonti04} and the position of \gal\ is marked by the
   blue cross (errorbars are 1 $\sigma$).}
  \label{fig:bpt}
\end{figure}

\begin{table}
  \begin{center}
  \begin{minipage}{174pt}
  \caption{UVOT photometry from pre-discovery data and the discovery observation.}\label{tab:uvotPhot}%
  \begin{tabular}{@{}ccc@{}}
  \hline
  Filter  & Magnitude         & Magnitude \\ 
          & (AB mag)          & (AB, discovery obs)\\
  \midrule
    v     & $15.78\pm0.07$    &  $15.82\pm 0.10$  \\
    b     & $16.36\pm0.07$    &  $16.41\pm 0.08$ \\
    u     & $17.40\pm0.07$    &  $17.41\pm 0.09$ \\
    uvw1  & $18.33\pm0.08$    &  $18.39\pm 0.10$  \\
    uvm2  & $18.95\pm0.08$    &  $18.88\pm 0.11$ \\
    uvw2  & $18.85\pm0.07$    &  $18.91\pm 0.09$ \\
  \hline
  \end{tabular}
  \end{minipage}
  \end{center}
\end{table}

\begin{table}
  \begin{center}
  \begin{minipage}{174pt}
  \caption{Liverpool Telescope upper limits on emission at the position of \swj\ after subtracting the
  galaxy emission (AB magnitudes).}\label{tab:ltPhot}%
  \begin{tabular}{@{}ccccc@{}}
  \hline
  MJD-59752      &  $g$     &  $r$     & $i$     & $z$   \\
  \midrule
  7.20           & $>$20.8  & $>$20.8  & $>$20.9 & --       \\ 
  32.20          & $>$21.7  & $>$21.6  & $>$21.7 & $>$21.5  \\  
  25.15          & $>$20.5  & $>$21.2  & $>$21.4 & $>$21.0  \\
  29.18          & $>$20.6  & $>$20.7  & $>$20.6 & $>$20.1  \\
  33.21          & $>$22.0  & $>$21.2  & $>$19.6 & $>$17.0  \\
  66.12          & $>$21.5  & $>$21.6  & $>$21.5 & --       \\
  \hline
  \end{tabular}
  \end{minipage}
  \end{center}
\end{table}

\clearpage



\section*{Supplementary Information}

\subsection*{Discussion}

\subsubsection*{The nature of the host galaxy, \gal}

A natural question is whether we have detected a new source, or are just seeing normal AGN activity.
The first identified QPE occurred in an AGN, the Seyfert-2 GSN 069, but the quasi-periodic nature of
the eruptions were not consistent with observed AGN activity \cite{Miniutti19}. \swj\ has a much
longer (quasi-)period than the previously identified QPEs, but its behaviour remains inconsistent
with typical AGN flaring. Rapid variability, such as that seen in the rise and decay of \swj\ is
sometimes seen in Seyfert galaxies, and recently a narrow-line Seyfert 1 galaxy was shown to undergo
a very soft outburst with a similar spectrum to \swj, and a decline similar in timescale to that
seen on day 4 of \swj\ \cite{Boller21}. However, in that object the spectrum hardened as the
luminosity declined -- the opposite behaviour to that seen in \swj. Further, our optical spectrum
(Figure~2), is inconsistent with a narrow-line Seyfert
classification: such galaxies show broader spectral lines and strong Fe{\sc ii} emission
\cite{Tarchi11}. The BPT diagram (Extended Data Figure.~3) further demonstrates that the optical
spectrum is not consistent with a Seyfert, and is only marginally consistent with a weak AGN.

Indeed, there are many objections to classifying \gal\ as a form of AGN. AGN are typically identified by their
hard (2--10 keV) X-ray flux, but \swj\ was never detected in this band. Summing up all of the 2--10
keV data, 148 ks in total, we obtain a marginal detection (3.2-\s\ significance, adopting the method
of \cite{Kraft91}). Assuming a power-law spectrum with a photon index of 1.7, this yields $L_{\rm
2-10} = (3.8^{+1.6}_{-1.4})\tim{40}$ \ergs. This does not alone rule out AGN activity, but does
require any AGN to be very weak, consistent with the result of the optical spectrum.
Furthermore, the host galaxy does not appear in the ALLWISEAGN catalogue \cite{Secrest15}, which
contains 84\%\ of all AGN brighter than \gal. It is present in the WISE catalogue, with colours
$W1-W2 = 0.14$, $W2-W3 = 3.9$. The $W1-W2$ colour is much more blue than the AGN shown in the
classification plot (fig.~12) of \cite{Wright10}. 

The difficulty of explaining the quasi-periodic nature, timescale and luminosity of the outbursts in
a weak, non-Seyfert AGN, combined with the above arguments against \gal\ as an AGN, lead us to rule
out variability of an existing AGN as a plausible explanation for \swj.

\subsubsection*{The association with SN2020rht}

While the \swift-XRT position for \swj\ is coincident with the nucleus of the galaxy \gal, we cannot rule out spatial
colocation with SN2020rht. However, as noted in the main article above, it is difficult to identify a mechanism
by which SN2020rht could evolve into \swj. Accretion is ruled out on simple energetics grounds: the Eddington luminosity
for a stellar-mass black hole is $\til10^{38}$ \ergs, and any theoretical super-Eddington (to the tune of 4 orders of magnitude!)
emission would not exhibit the soft, thermal spectrum we observed -- and would show strong optical emission along with the X-rays.
Magnetar spin-down has been proposed as powering long-lived GRB emission (for example, refs~\cite{Gompertz13}) and can therefore
certainly provide sufficient energy; however, existing models are not consistent with \swj, as already detailed above.

We thus regard the near-alignment with SN2020rht as simply a chance occurrence. A detailed calculation of the probability of this
is beyond the scope of this paper, especially as the rate of X-ray transients in the Universe is not a well-known quantity; however,
a simple consideration is enough to show that such alignments cannot be rare. If the typical supernova rate is
0.01--0.1 per galaxy per year \cite{Anderson13}, and given that the XRT localisation accuracy is comparible to or
larger than the angular size of a typical galaxy, it follows that the probability of an XRT transient in a given galaxy
also lying close to a supernova of up to 2 years old, must be of order a few percent to a few tens of percent.

\medskip
\subsection*{Comparison of \swj\ with similar objects}
\label{compa}
\smallskip

There is a growing collection of systems identified as QPEs, with varying degrees of confidence (and
cycles observed) \cite{Miniutti19,Giustini20,Song20,Arcodia21,Chakraborty21}. These systems share
various common properties. They have all been discovered in X-rays and show little (if any)
variability at longer wavelengths. They are all located in the nucleus of their host galaxies, and
have very soft spectra, with almost all emission below 2~keV and capable of being fitted by an
absorbed blackbody with temperatures in the \til100--200 eV range. They all exhibit a clear
correlation between spectral hardness and luminosity, and have peak luminosities of $L_{\rm 0.3-2
keV} \approx 10^{42-43}$ \ergs. They show outbursts that are approximately, but not strictly
periodic. They are typically reported to have black hole masses of $\til10^5$ \msol. These
properties are all shared by \swj, with the exception of the timescales. The QPE candidates are all
found to have outbursts with durations of minutes to hours, and periods of a few hours; \swj\ as
noted has a much longer outburst duration and period.

A smaller number of objects have been classified as PNTs \cite{Payne21,Payne22,Wevers22}. These
objects were discovered optically and show much longer periods ($>100$ d) but are again repeating
events colocated with a galaxy nucleus, this time with black hole masses $\til10^8$ \msol. In the
first of these, ASASSN-14ko \cite{Payne21,Payne22}, no X-ray outburst is seen although the X-ray
emission from the host AGN is found to decrease prior to the optical outburst. The second event,
AT2018fyk \cite{Wevers22}, does show a strong X-ray and UV outburst which was abruptly terminated,
before rising again about 600d later.

A further system, eRASSt J045650.3-203750 \cite{Liu23}, shows similarities to both classes of object.
It was discovered via its X-ray outbursts, which are \til90 d in duration and occur every \til220 d.
Spectrally, this object is harder than the QPEs, with notable flux detections above 2 keV. It is again
located coincident with a galactic nucleus, with an estimated SMBH mass of $\til10^6$ \msol. In this event,
some evidence for UV variability is seen but no optical variation.

All of these objects have been interpreted by the papers cited above as rpTDE events, with differing black-hole masses
and donor types. \swj\ is clearly most similar to the QPEs and eRASSt J045650.3-203750, but with an outburst duration
and period much longer than the former, and shorter than the latter.

\subsubsection*{The nature of \swj}
\label{sec:natureSWJ}

Here, we discuss the properties of \swj\ and estimate physical quantities such as the mass accreted
during each outburst. Given the discussion above, we assume that the variability is not caused by
activity within a standard AGN disc (by, for example, the mechanisms outlined in
\cite{Cannizzaro20,Sniegowska20}). We therefore explore the implications in the context of
existing models for accretion of material tidally stripped from stars orbiting near the central
black hole \cite{Zalamea10,King20,Metzger22,Lu23}.

If the outbursts are accretion events then the total mass accreted during an outburst is given by:

\begin{equation}
  M_{\rm acc} \til \frac{E_{\rm ob} k}{\eta c^2}
\end{equation}

\noindent where $E_{\rm ob}$ is the measured X-ray energy during the outburst, $k$ is a bolometric
correction, and $\eta$ is the radiative efficiency of the accretion process. We can roughly
characterise the observed outbursts as lasting $\approx$10 days, with a mean X-ray luminosity of $L_{\rm 0.3-2}
\til 3\tim{42}$ \ergs, thus $E_{\rm ob} \til 3\tim{48}$ erg, hence $M_{\rm acc} \approx
3\tim{27} k/\eta$ g. A typical value for the accretion efficiency is  $\eta\til0.1$ \cite{Shakura73}. The lack of observed
emission in the UV or optical is consistent with the simple blackbody model fitted to the XRT
spectrum, from which we find $k\til1$. A correction is needed to account for the absorption,
however this change is found (via fitting in {\sc xspec}) to be very small; therefore, $M_{\rm acc}
\approx 10^{-5} \msol$ is a reasonable estimate of the mass accreted during an outburst.

We can provide an estimate of the black hole mass in \swj\ by comparing the measured temperature of
$\til 100$\,eV ($1.16\times 10^6\,$K) with the peak temperature of a standard accretion disc
\cite{Shakura73}. The temperature profile of a standard disc accreting at a rate ${\dot M}$ is 

\begin{equation}
T_{\rm eff}(R) = \left\{\frac{3GM{\dot M}}{8\pi R^3\sigma}\left[1-\left(\frac{R_{\rm in}}{R}\right)^{1/2}\right]\right\}^{1/4}\,,
\end{equation}

\noindent and thus the maximum temperature reached is

\begin{equation}
T_{\rm max} = \frac{6^{6/4}}{7^{7/4}}\left(\frac{3GM{\dot M}}{8\pi \zeta^3 R_{\rm g}^3 \sigma}\right)^{1/4} 
            = \frac{6^{6/4}}{7^{7/4}}\left(\frac{3c^6{\dot M}}{8\pi \zeta^3\sigma G^2 M^2}\right)^{1/4} \label{tmax}
\end{equation}

\noindent where the innermost stable circular orbit occurs at $R_{\rm isco} = \zeta R_{\rm g}$
(with $R_{\rm g} = GM/c^2$ and $\zeta$ ranging from unity for a maximally spinning prograde black
hole to $9$ for a maximally spinning retrograde black hole), and the numerical pre-factor of
$6^{6/4}/7^{7/4} \approx 0.488$ is due to the zero-torque inner boundary condition (see
\cite{Nixon21} and references therein for a discussion of discs with non-zero torque boundary
conditions).

Taking the total mass accreted during a typical outburst (derived above) of $10^{-5}$\ \msol\ over a
period of 10 d, yields an accretion rate of $\sim 2\times 10^{22}$ g s$^{-1}$. Putting this into
Equation~\ref{tmax} gives an estimate for the black hole mass of $M_{\rm est} \til 2.4\times
10^{5}\,(k^{1/2}/\zeta^{3/2}) \msol$, similar to those reported for QPE sources
\cite{Miniutti19,Giustini20,Arcodia21,Song20}. We note that for the blackbody temperature range
of 50--250 eV (Extended Data Figure.~2), the black hole mass estimate ranges from 6\tim{4}\ \msol\ to 1.5\tim{6}\ \msol.
With an estimate of the black hole mass, we can now consider the types of system which can produce 
an rpTDE consistent with the behaviour of \swj.

As discussed in the main paper, the formation of a system in which rpTDE can take place requires the
Hills mechanism \cite{Hills98}, in which a supermassive black hole disrupts a stellar
binary system, resulting in one star being ejected and the other being bound to the black hole. In
some cases the bound star may have a sufficiently small pericentre distance that it is (fully or
partially) disrupted by the black hole, and \cite{Cufari22} propose this as a route for forming
rpTDEs. They show that the orbital period of the bound star is

\begin{equation}
\label{porb}
P_\bullet \simeq \pi \frac{a_\star^{3/2}}{\sqrt{2GM_\star}}\left(\frac{M_\bullet}{M_\star}\right)^{1/2}\,,
\end{equation}

\noindent where $M_\bullet$ is the mass of the black hole, $M_\star$ the mass of the primary star.
The semi-major axis of the (pre-disruption) binary star system, $a_\star$, is strongly constrained.
For $a_\star \gtrsim GM_\star/\sigma^2 \til 0.02$ AU (for $M_\star = \msol$ and stellar velocity
dispersion $\sigma = 200 $\ km s$^{-1}$), the binary system would be destroyed by the tidal field of
the galaxy centre before reaching the central black hole, while for $a_\star \lesssim 0.001$ AU it
is hard to fit a main-sequence star of any mass inside the binary orbit. If the progenitor of \swj\
were a binary consisting of a compact object (e.g.\ a white dwarf) and a low-mass main-sequence star,
with a semi-major axis of $a_\star =$ 0.005--0.01 AU, and the compact object was ejected during the
Hills encounter, then for the low-mass star to enter into a bound orbit around the black hole with
the observed \til25 d period requires a black hole mass of $M_\bullet \sim 4\times 10^5\,\msol$.
This is consistent with the mass estimate derived from the temperature of the X-ray spectrum. It is
worth noting that the exact value of the orbital period of the bound star depends somewhat on the
details of the Hills capture process. For example, it is possible to get periods of order \til25 d
with a larger black hole mass (of order $10^7\,M_\odot$), but as shown in the probability distribution
functions (PDFs) displayed in fig.~1 of \cite{Cufari22} such cases are relatively unlikely with the
PDF of the orbital period strongly peaked around the value given by equation \ref{porb}.

The pericentre distance from the black hole at which the star can be partially disrupted is $\approx
1.6(M_\bullet/M_\star)^{1/3}R_\star$ (for example, refs~\cite{Coughlin22}) where $R_\star$ is the radius of
star. The pericentre distance of the stellar orbit must therefore be $r_{\rm p} \gtrsim 50 R_{\rm
g}$, where $R_{\rm g} = GM/c^2$ is the gravitational radius of the black hole (assuming a low-mass
star and a black hole of mass a few$\times 10^5\,M_\odot$). This suggests why the outburst period of
\swj\ is so long compared to the other QPE sources with a similar black hole mass. The QPE sources,
with white-dwarf donors, must have pericentre distances that are $\lesssim 5 R_{\rm g}$ to liberate
any mass from the white dwarf (indeed in many cases the orbits calculated by \cite{King22} have
pericentre distances that imply the black hole must be rapidly spinning in the prograde direction to
avoid directly capturing the star in a single orbit). Thus, in those sources, any accretion flow
that forms will evolve rapidly due to the small circularization radius of the stellar debris,
whereas here we have a much larger circularization radius of $\til 50-100 R_{\rm g}$. The viscous
timescale for a standard disc with $M_\bullet \til 3\tim{5} $ \msol, $\alpha \approx 0.3$ (see
ref~\cite{Martin19}) and disc angular semi-thickness, $H/R \til 0.1$, is of the order of hours for radii
of order a few $R_{\rm g}$, and grows to of the order of days for radii of order $50 R_{\rm g}$.
This provides the correct timescales observed for both the QPE sources and \swj. It is therefore
possible to explain \swj\ as the repeated partial tidal disruption of a solar-type star by a
modest-mass black hole, within the framework of a model that can also explain the shorter-period
QPEs (white dwarfs orbiting modest-mass black holes) and the longer-period PNTs (main sequence stars
around more massive black holes).

Our black-hole mass estimate above was based on the X-ray spectrum as in previous QPE papers
(for example, ref~\cite{Miniutti19}); however, other methods may also be considered. If we assume that the
peak accretion rate is at the Eddington limit then this requires a black hole mass of $1.6\tim{5}$
\msol, although this is probably a lower limit as the accretion is more likely to be sub-Eddington.
Using the {\sc starlight} fits to the optical spectrum (Figure. 2) we can estimate the total stellar
mass in the host galaxy \gal\ as \hbox{$(2.93 \pm 0.94)\tim{9}$~\msol}. The relationship between
stellar mass and bulge mass shown by \cite{Reines15} (see their fig.~8) and \cite{Schutte19} (fig.
4) shows that the expected black hole mass is in the range $\log M_{\rm BH}/\msol \approx$5--7,
consistent with the above measurements. Recently, \cite{Wevers22b} obtained spectroscopic
observations to determine the black hole masses of various QPE hosts using velocity dispersion,
finding low masses similar to those predicted by the X-ray temperature approach. This may indicate
that QPEs preferentially occur in galaxies with unusually low black hole masses; spectroscopic
measuresments of \gal\ in order to measure the velocity dispersion will be needed to confirm the
black hole mass in this system.


Another interpretation for QPEs that is worth mentioning here is put forward by (see also
refs~\cite{Metzger17b,Metzger22}). In this model there are two stars inspiralling towards the black
hole due to gravitational wave losses. As the more massive star emits stronger gravitational waves,
its orbit shrinks faster such that it can catch up to a lower mass star. This can lead to strong
interactions between the stars, stripping gas that can then periodically fuel the accretion flow on
to the central black hole. For stars that are orbiting in the same plane but with opposite
directions (i.e. one prograde and one retrograde) the stars encounter each other once per orbit,
whereas for stars orbiting in the same direction the encounter timescale is many orbits.
\cite{Metzger22} predict that the former results in periodicity of order hours and the latter of
days. Further \cite{Metzger17b} analysed non-coplanar orbits and predicted recurrence times of years
to centuries. Thus, this model is capable of explaining the range of timescales from QPEs, to \swj,
to PNTs and beyond. To explain the $\sim 25$\,d variability of \swj\ a coplanar set of stars is
required. As the stars are likely to encounter the black hole on randomly distributed orientations,
a mechanism is required to ensure the stars settle into a coplanar state. \cite{Metzger22} appeal to
Type I inward migration through a gaseous AGN disc, which in the case of \swj\ may not be present.

\paragraph{Other possible interpretations}

Other possible explanations for \swj\ may be considered, although we find none of them as compelling
as the rpTDE model described above. Repeating stellar phenomena such as X-ray binary outbursts can
be disregarded on luminosity grounds. Magnetar-powered emission can produce the required luminosity
(for example, refs~\cite{Gompertz13,Metzger14,Margalit18}). We have above discussed and discarded the
possibility of a magnetar connected to the two-year old SN2020rht; more generally, while a magnetar
could produce emission with luminosity and spectrum consistent with \swj, the variability observed
cannot be explained by this model. The spectral variation observed rules out the possibility of a
steadily-emitting magnetar periodically obscured by some absorbing disk or stream, and the absence
of strict periodicity rules out binary eclipses.

A class of objects known as `Fast X-ray transients' has been identified in archival data
(for example, refs~\cite{Jonker13,Bauer17,Lin22}), which undergo relatively rapid X-ray outbursts. However, these are
spectrally harder and more luminous than \swj, and have much shorter outbursts ($<50$ ks
\cite{Quirola22}). Further, they have not been observed to repeat; indeed \cite{Quirola22} only
identify as FXT candidates those objects which were only detected once.

A more promising type of analogous object is HLX-1 in ESO 243-49, which has been proposed as an
accreting intermediate mass black hole (IMBH; a black-hole with a mass of $10^{3-4}$ \msol)
\cite{Farrell09}. This system has a similar peak luminosity to \swj, but a harder spectrum and much
longer period. Various models have been proposed to explain this system, including wind accretion
\cite{Miller14} and disc instabilty \cite{Soria17}, or rpTDE akin to that proposed above
\cite{King22}. As shown above, simple energetics suggests that the amount of matter accreted during
a single outburst (and thus stripped from the star every \til25 d) is $\til10^{-5} \msol$, orders of
magnitude too high to be powered by a stellar wind. In the disc-instability model \cite{Soria17},
the accretion disc is formed by Roche-Lobe overflow, which can provide the requisite amount of mass.
In their preferred scenario, the accretion is modulated by a disc wind instability. Futhermore, the
HLX outbursts show a fast-rise, slow-decay morphology which is the opposite to that seen for \swj.
The differing shape, much shorter period, lack of optical variability, and softer spectra than HLX-1
all pose significant challenges for this interpretation. We therefore conclude that the progenitors
of HLX-1 and Swift J0230 are likely to be physically distinct.

We thus conclude that rpTDE is the most likely explanation for \swj, although detailed numerical modelling of the accretion
flows is required to confirm whether the deviations from strict periodicity and occasional long quiescent times
can be explained by this model.

\section*{Declarations}

\subsection*{Data Availability}

All of the \emph{Swift} data are available via the Swift data archives provided in the USA
(\url{https://swift.gsfc.nasa.gov/archive/}), UK (\url{https://www.swift.ac.uk/archive/}) and Italy
(\url{https://www.ssdc.asi.it/mmia/index.php?mission=swiftmastr}); they have targetIDs 00014936 and
00015231. Reduced \swift-XRT data for this transient are available at
\url{https://www.swift.ac.uk/LSXPS/transients/690}. The \emph{Chandra} data are publicly available
via the \emph{Chandra} data archive (\url{https://cxc.harvard.edu/cda/}), with sequence 704871 and
obsID 27470. The NOT data will be available through the NOT public interface after the expiration of
the standard proprietary period; the reduced spectrum is available through the University of
Leicester FigShare repository (\url{https://doi.org/10.25392/leicester.data.c.6444296}). The
Liverpool Telescope Data will be available through the Liverpool Telescope public interface after
the expiration of the standard proprietary period; the photometry was included in this published
article.

\section*{Acknowledgments}

This work made use of data supplied by the UK Swift Science Data Centre at the University of
Leicester.
We acknowledge the following funding support: UK Space Agency, grant ST/X001881/1
(PAE, KLP, RAJE-F and AAB). 
The Science and Technology Facilities Council, grants ST/Y000544/1 (CJN), and ST/W000857/1 (POB).
The Leverhulme Trust, grant RPG-2021-380 (CJN).
Italian Space Agency, contract ASI/INAF n. I/004/11/5 (SC).
European Union’s Horizon 2020 Programme under the AHEAD2020 project, grant 871158 (RAJE-F).
European Research Council under the European Union’s Horizon 2020 research and innovation programme,
grant 725246 (DBM).
The Cosmic Dawn Center (DAWN) is funded by the Danish National Research Foundation under grant No. 140.
The Pan-STARRS1 Surveys (PS1) and the PS1 public science archive have been made possible through
contributions by the Institute for Astronomy, the University of Hawaii, the Pan-STARRS Project
Office, the Max-Planck Society and its participating institutes, the Max Planck Institute for
Astronomy, Heidelberg and the Max Planck Institute for Extraterrestrial Physics, Garching, The Johns
Hopkins University, Durham University, the University of Edinburgh, the Queen's University Belfast,
the Harvard-Smithsonian Center for Astrophysics, the Las Cumbres Observatory Global Telescope
Network Incorporated, the National Central University of Taiwan, the Space Telescope Science
Institute, the National Aeronautics and Space Administration under Grant No. NNX08AR22G issued
through the Planetary Science Division of the NASA Science Mission Directorate, the National Science
Foundation Grant No. AST-1238877, the University of Maryland, Eotvos Lorand University (ELTE), the
Los Alamos National Laboratory, and the Gordon and Betty Moore Foundation. LI was supported by
grants from VILLUM FONDEN (project number 16599 and 25501) We thank Andy Beardmore for help with the
bootstrapping method the period analysis.

\subsection*{Author contributions}

PAE authored the tools that discovered the event, was PI of the \swift\ and \emph{Chandra} observations, 
performed most of the X-ray data analysis and led the writing of the article. CJN carried out the theoretical
interpretation of the data, and produced the associated text. SC first noticed the automated report 
of the new transient and classified it as of interest; he was CoI of the \emph{Chandra} observations.
PC obtained the NOT spectrum and led the analysis of it, producing Figure 2 and Extended Data Figure.~\ref{fig:bpt}.
DAP obtained and analysed the Liverpool Telescope data. AAB led the analysis of the UVOT data.
KLP carried out some of the XRT data analysis, particularly spectral fitting. SRO supported the UVOT data
analysis. RAJE-F was a CoI of the \emph{Chandra} observations and was involved in many discussions concerning
the interpretation of the object. DBM arranged for the acquisition of the NOT spectrum and helped with
its analysis and interpretation, and was a CoI \emph{Chandra} observations. LI conducted the analysis of the NOT
spectrum with {\sc starlight}. MRG and PTO offered AGN expertise
supporting ruling out AGN activity as the cause of the observed outburst. JPO and BS offered programmatic support
and general input. All authors read the text and contributed to its editing.

\subsection*{Competing interests}

The authors declare no competing interests.

\bibliographystyle{naturemag}

\bibliography{phil}

\end{document}